# West-WRF AI 2-km: High-Resolution Prediction of Integrated Vapor Transport and Precipitation

Nazak Rouzegari,[a] Vesta Afzali Gorooh,[b] Agniv Sengupta,[b] Phu Nguyen,[a] Kuo-Lin Hsu,[a] Amir AghaKouchak,[a] Soroosh Sorooshian,[a] F. Martin Ralph,[b] and Luca Delle Monache[b]

[a]*Center for Hydrometeorology and Remote Sensing (CHRS), The Henry Samueli School of Engineering, Department of Civil and Environmental Engineering, University of California, Irvine, USA*
[b]*Center for Western Weather and Water Extremes (CW3E), Scripps Institution of Oceanography, University of California, San Diego, USA*

ABSTRACT: We introduce a stretched-grid artificial intelligence (AI) weather forecasting model with 2-km resolution over the western United States and part of the Northeast Pacific and approximately 31-km resolution elsewhere globally. Forecasting over the western U.S. is challenging because complex topography and atmospheric rivers (ARs) strongly influence orographic precipitation. West-WRF AI 2-km builds on a global model pretrained with a 40-year European Centre for Medium-Range Weather Forecasts Reanalysis v5 (ERA5) dataset and is fine-tuned with the Center for Western Weather and Water Extremes (CW3E) 2-km regional reanalysis to produce autoregressive 6-hourly forecasts of precipitation and integrated vapor transport (IVT). Forecasts are evaluated over winters 2020–2023 using gridded precipitation observations, rain gauges, and AR Reconnaissance dropsondes and are benchmarked against coarser-resolution AI forecasts and regional and global numerical weather prediction (NWP) systems. West-WRF AI 2-km reproduces observed precipitation-intensity distributions, retains fine-scale spectral variability, and produces sharper narrow coastal precipitation bands and localized, terrain-sensitive extremes. Its broader-scale performance remains comparable to coarser-resolution configurations while preserving large-scale skill despite higher resolution. Dropsonde verification shows lower errors and improved categorical skill at the most extreme IVT threshold. Overall, West-WRF AI 2-km provides its greatest value for localized precipitation extremes and intense AR-related moisture transport.

---

*Corresponding author:* Nazak Rouzegari, nrouzega@uci.edu

## 1. Introduction

Across the western United States (U.S.), precipitation is shaped by interactions among large-scale atmospheric circulation, steep terrain, and strong coastal-to-inland hydroclimatic gradients. These controls produce sharp spatial contrasts in rainfall and snowfall, making accurate precipitation predictions especially challenging in regions where water supply, flood risk, and ecosystems are highly sensitive to individual storms and seasonal variability. Along the western U.S. coast, large-scale circulation interacts with complex topography to further modulate precipitation through orographic processes (Hsieh et al. 2025). Simulating orographic precipitation is difficult because it involves processes spanning large-scale moisture transport, mesoscale turbulence, and cloud microphysics (Smith 2019). Therefore, coarse-resolution models frequently struggle to capture precipitation variability because they inadequately represent topography, small-scale atmospheric processes, and localized convective dynamics.

Winter precipitation is especially important along the western U.S. coast, which receives about 50–75% of its annual precipitation from November through March, much of which is associated with atmospheric rivers (ARs) (Dettinger et al. 2011; Hecht and Cordeira 2017; Hsieh et al. 2025; Sengupta et al. 2025). ARs are narrow, elongated corridors that are typically about 2,000 km long and 400–800 km wide and extend up to 3 km above the surface (Cobb et al. 2023; Ralph et al. 2020). They transport substantial moisture and energy from lower to higher latitudes (Lu et al. 2025; Scholz and Lora 2024) and can produce intense precipitation when interacting with coastal terrain (Lu et al. 2025). ARs, which account for many extreme precipitation events in mid-latitude regions, especially along the U.S. West Coast, are expected to become stronger as the climate warms (Higgins et al. 2025; Lu et al. 2025; Webb et al. 2026). Strong or persistent ARs can cause destructive flooding and damaging winds, resulting in substantial loss of life and damage to critical infrastructure (Corringham et al. 2022; Higgins et al. 2025; Webb et al. 2026). Accurate forecasts are therefore essential for weather- and water-related decision-making across the region (Lam et al. 2023).

Traditionally, precipitation forecasting has relied on numerical weather prediction (NWP) systems, which solve the governing atmospheric equations and use parameterizations to represent unresolved processes (Baño-Medina et al. 2025; Lam et al. 2023). Examples include the Global Forecast System (GFS; NCEP 2020a) and the Global Ensemble Forecast System (GEFS; NCEP 2020b), the European Centre for Medium-Range Weather Forecasts (ECMWF) Integrated

Forecasting System (IFS; ECMWF 2022), the High-Resolution Rapid Refresh (HRRR) (Dowell et al. 2022), and the Center for Western Weather and Water Extremes (CW3E) Western U.S. Weather Research and Forecasting (West-WRF) 200-member ensemble (Delle Monache et al. 2025).

Although high-resolution regional NWP models can better represent fine-scale processes than coarser-resolution global NWP models, they remain computationally expensive and dependent on parameterizations. Moreover, coarse global models may underestimate extreme precipitation, particularly during ARs (Baño-Medina et al. 2025).

These limitations have increased interest in artificial intelligence (AI)-based weather prediction models that can generate forecasts more rapidly and with lower computational and energy costs (Lang et al. 2024). Examples include Artificial Intelligence Forecasting System (AIFS) (Lang et al. 2024), AIFS ENS (Lang et al. 2026), GraphCast (Lam et al. 2023), GenCast (Price et al. 2025), CorrDiff (Mardani et al. 2025), FourCastNet (Pathak et al. 2022), Pangu-Weather (Bi et al. 2023), MetNet-3 (Andrychowicz et al. 2023), and StormCast (Pathak et al. 2026). Despite their skill, the relatively coarse resolution of global AI models can limit their representation of extremes, complex terrain, and regional AR-related precipitation, particularly over the western U.S. (Baño-Medina et al. 2025).

This has motivated the development of regional AI weather forecasting models with improved local-scale representation. Limited-area models (LAMs) restrict forecasting to a regional domain (Xu et al. 2025). In contrast, stretched-grid models increase resolution over a target region while retaining coarser global coverage (Lang et al. 2024; Nipen et al. 2025). The stretched-grid approach is particularly suitable for AR prediction because it preserves the large-scale atmospheric context governing moisture transport while resolving local terrain and mesoscale processes. ECMWF's Anemoi framework supports the development of such systems through flexible graph-based architectures, customized grids, and user-defined datasets (Lentze 2024).

Building on the previous CW3E West-WRF AI 6-km framework, this study develops West-WRF AI 2-km, a deterministic, stretched-grid AI-based weather forecasting system using the Anemoi framework. The model provides approximately 2-km resolution over the western U.S. and part of the Northeast Pacific while maintaining approximately 31-km resolution elsewhere globally. This configuration combines global atmospheric context with enhanced regional detail to improve medium-range forecasts of surface and pressure-level hydroclimatological variables, with a particular focus on precipitation and integrated vapor transport (IVT) during AR events. Unlike recent regional AI systems based on limited-area or stretched-grid approaches (Adamov et al. 2025; Nipen et al. 2025), West-WRF AI 2-km focuses specifically on AR-related IVT and orographic precipitation over the western U.S. To the best of our knowledge, West-WRF AI 2-km is the first AI weather forecasting system evaluated specifically for AR prediction at approximately 2-km resolution over the region.

## 2. Methodology and Data Sources

### *a. Training Datasets*

*CW3E 2-km Reanalysis:* The CW3E 2-km reanalysis, produced with WRF v4.2.2 over western North America and the near-coast eastern North Pacific and forced by ERA5, with annual ERA5-Land and sea-surface-temperature initialization (Hersbach et al. 2020; Worsfold et al. 2024), serves as the training reference for the regional model. The 6-hourly, nested 2-km dataset spans 2012–2023 and is designed to resolve ARs, complex terrain, and extreme precipitation over the western U.S.

*ERA5 31-km reanalysis:* Global pretraining uses approximately 40 years of ERA5 reanalysis (Hersbach et al. 2020) on the N320 reduced Gaussian grid (Baño-Medina et al. 2025); because ERA5 assimilates observations unavailable in real time, it may provide more accurate initial states than operational analyses, a difference to consider when comparing with operational dynamical systems. Regional fine-tuning uses ERA5 and CW3E reanalysis fields from 2012–2020 (eight years), with winters 2020–2021 through 2022–2023 reserved for independent evaluation.

The model uses prognostic, diagnostic, and forcing variables from both the CW3E 2-km reanalysis and global ERA5 (N320 grid), consistent with the West-WRF AI 6-km model (Baño-Medina et al. 2025; Hsieh et al. 2025); full variable definitions, vertical levels, and forcing fields are listed in Table S1. This study focuses on total precipitation and IVT, with particular emphasis on extremes, evaluated over three winters (November–March 2020–2023) against the benchmark models described below.

### *b. Training Framework*

Fig. 1 shows the overall architecture, based on a multi-stage training strategy implemented within the ECMWF Anemoi Graph Transformer framework, using a stretched-grid configuration for approximately 2-km resolution over the western U.S. within an approximately 31-km global grid. Training proceeds through three sequential stages (Baño-Medina et al. 2025). In Stage A1, the global model is pretrained at approximately 31-km resolution using 40 years of ERA5 N320 data with a one-step, 6-hour rollout, learning large-scale atmospheric dynamics; Baño-Medina et al. (2025) then transferred these representations to a 6-km stretched-grid configuration using the CW3E 6-km reanalysis, which this study extends to 2 km. In Stage B1, the Stage A1 weights are fine-tuned using the CW3E 2-km reanalysis to refine the regional grid while retaining the 31-km global background. Stage B2 increases the training rollout from one step to eight autoregressive steps (48 hours; one additional step per epoch, using the same Stage B1 dataset) to improve temporal stability. During inference, the trained model is iterated autoregressively for up to 24 six-hour steps (144 hours). All stages minimize a variable- and level-weighted RMSE loss, with the regional domain scaled to account for 25% of the total loss.

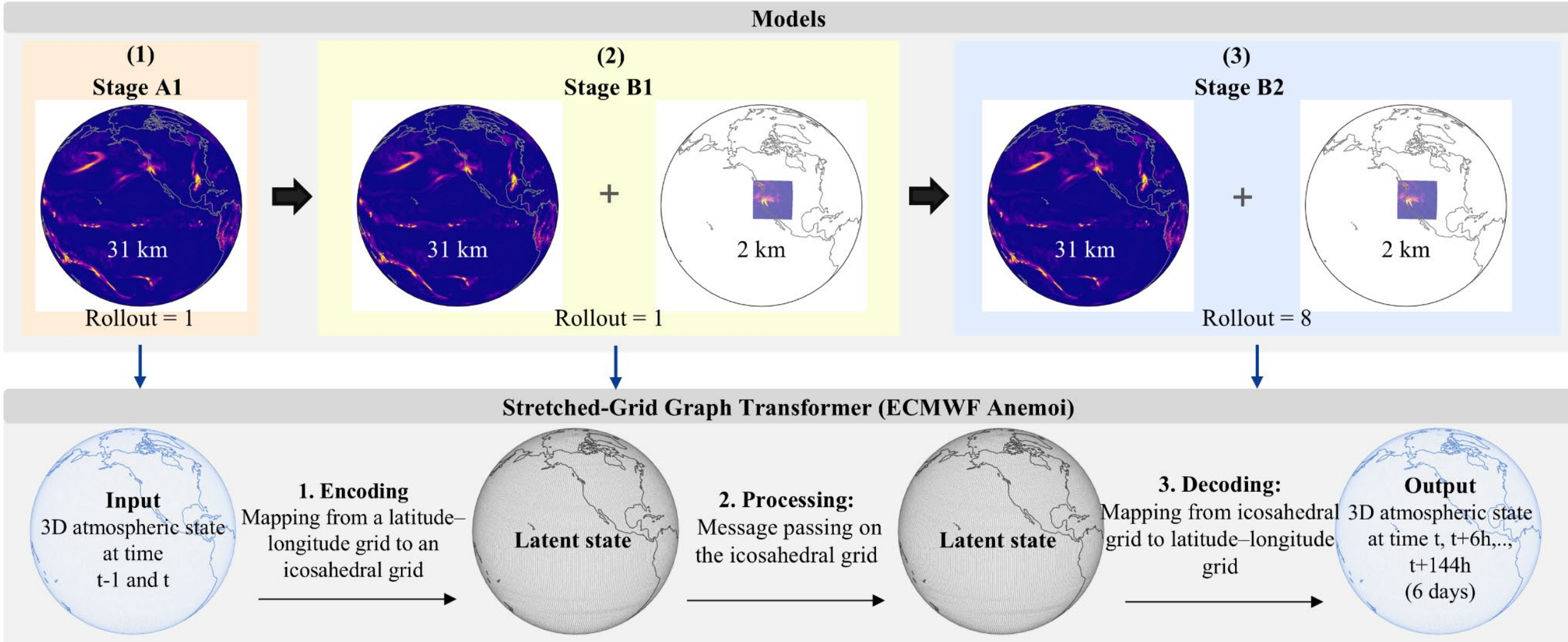

**Fig. 1.** Schematic overview of the West-WRF AI 2-km data-driven forecasting system. The upper panel illustrates the three-stage training and inference workflow. Stage A1 uses the 40-year global 31-km atmospheric state with a one-step rollout. Stage B1 combines the 31-km global state with a nested 2-km regional state, also using a one-step rollout. Stage B2 uses an eight-step (48 hours) autoregressive rollout during training. During inference, the trained model is iterated autoregressively for up to 24 six-hour steps (144 hours). The lower panel shows the stretched-grid Graph Transformer architecture, based on European Centre for Medium-Range Weather Forecasts (ECMWF) Anemoi, that is used in each of the three stages to process the corresponding global or coupled global–regional atmospheric states. Atmospheric fields on a latitude–longitude grid are encoded onto an icosahedral latent grid, processed through graph-based message passing, and decoded back to the latitude–longitude grid to produce 6-hourly atmospheric forecasts from initialization time *t* through *t + 144* hours.

Empirical sensitivity analyses of Stage B1 hyperparameters, as well as the selected hyperparameter settings used for Stages B1 and B2, are presented in Figs. S1–S3 and Table S2 of the Supplemental Material.

### c. *Stretched-grid Graph Transformer*

The forecasting architecture is the ECMWF Anemoi Graph Transformer (Lang et al. 2024), in which spatial locations and their interactions are represented as nodes and edges. The model takes 172 input channels (86 variables each from the current and previous time step; the diagnostic precipitation variable is treated separately), encoded onto an icosahedral graph at a global refinement level of 7 and a regional refinement level of 9 over the western U.S. (approximately 2.07 million grid points, 194,220 mesh nodes, and 45.0 million edges for both Stages B1 and B2). The encoder maps the input fields to a latent representation on the processor mesh (each mesh node connected to its 12 nearest input grid points); within the 512-channel processor, 16 Graph Transformer layers perform message-passing to model spatial interactions and atmospheric transport (Keisler 2022) while advancing the latent state one forecast step forward. The decoder then maps the latent representation back to the latitude–longitude grid (each output point connected to its 3 nearest processor mesh nodes), and this process is repeated autoregressively to produce forecasts to t+144 hours. Additional architectural details are given in Table S2 and in Nipen et al. (2025).

### d. *Computational Resources*

Training (Stages B1 and B2) takes approximately 9 days on 64 H100 graphics processing units (GPUs) (96 GB each), sharded 4 per node with data parallelism across 16 nodes (global batch size 16). Once trained, the model generates a complete 6-day forecast in approximately 3–4 minutes on a single H100 GPU, including data loading, preprocessing, and autoregressive generation through 144 hours (Table S2).

### *e. Observational Datasets*

*PRISM Precipitation:* Daily precipitation from the 4-km and 800-m PRISM Climate Group dataset (Daly et al. 2017, 1994) serves as the gridded observational reference. Forecasts are regridded to PRISM resolution and accumulated into 24-hour totals matching PRISM's −12 to +12 UTC window; because of the resulting 12-hour offset from the 00 UTC model initializations, a 1-day lead time corresponds to a +12- to +36-hour forecast window, with subsequent lead times adjusted accordingly.

*Gauge Precipitation:* Hourly gauge observations from the California-Nevada River Forecast Center's (CNRFC) quality-controlled precipitation estimate (QPE) product provide 526 valid stations across the common model domain (Fig. S4) for November–March 2020–2023. Observations are aggregated into right-labeled 6-hour accumulations (matching forecast temporal resolution) requiring all six hourly values; only initialization dates with gauge data at all lead times from 6–120 hours are retained. For each station, forecasts are extracted from the nearest model grid cell and paired with observations at matching valid times, combined across all three winters into a unified verification dataset.

*AR Reconnaissance Dropsondes IVT:* IVT derived from approximately 2,360 dropsonde profiles collected during 2021–2023 AR Reconnaissance missions (Ralph et al. (2020); 942 in Jan–Mar 2021, 637 in Jan–Mar 2022, and 781 in Nov–Dec 2022/Jan–Mar 2023) provides an independent reference computed directly from observed moisture and wind profiles rather than inferred from model-based reanalysis, reducing dependence on and potential shared biases with the forecast systems themselves.

### *f. Benchmark Models*

*West-WRF AI 6-km Precipitation and IVT:* West-WRF AI 2-km is compared with three benchmarks. West-WRF AI 6-km (rollout 8) is CW3E's predecessor data-driven regional system (Baño-Medina et al. 2025) — a stretched-grid Graph Transformer architecture trained on ERA5 that provides ~6-km forecasts of precipitation, IVT, and other atmospheric variables — and serves as the primary reference for quantifying the benefit of further resolution refinement.

*IFS 9-km Precipitation:* IFS 9-km, the physics-based ECMWF Integrated Forecasting System (2022), provides a global NWP benchmark. It is run four times daily and provides deterministic precipitation and IVT forecasts at approximately 9-km resolution (Tuppi et al. 2023).

*West-WRF 3-km Precipitation and West-WRF 9-km IVT:* West-WRF 3-km/9-km, CW3E's dynamical West-WRF system (Martin et al. 2018), provides a regional NWP benchmark via two deterministic simulations forced by the ECMWF high-resolution forecast (HRES). We use 6-hourly IVT at 9-km over the North Pacific and western U.S., and 6-hourly precipitation at 3-km over California and much of the West Coast.

The geographic domains for winter 2022–2023, including the West-WRF AI 2-km domain nested within the larger West-WRF AI 6-km and West-WRF 9-km domains, together with the gauge and dropsonde locations, are shown in Fig. S4. Because the benchmark domains differ in extent, only overlapping regions are used for intercomparison. Corresponding domains and dropsonde locations for 2020–2021 and 2021–2022 are shown in Figs. S5–S6.

### g. *Model Evaluation*

Precipitation forecasts are verified against gridded PRISM (regridded to a common grid) and against station-wise rain gauges, which capture localized features that may be smoothed in the gridded product; IVT is verified independently against AR Reconnaissance dropsondes, matched to the nearest model grid point and to the nearest valid forecast time within a ±3-hour window (dropsonde drift during descent is not considered (Cobb et al. 2021)). Evaluation covers lead times up to 120 hours (5 days). Because the West-WRF AI 2-km domain is nested within a course 31-km global grid, both the 2-km and surrounding 31-km regions are included when comparing against dropsondes.

#### 1) AR-BASED PRECIPITATION EVENTS:

AR-related events are identified and ranked using the AR scale of Ralph et al. (2019), using the same events (rank ≥ 2, impacting the San Francisco region, 37.5°N/122.5°W) listed in Table 1 of Baño-Medina et al. (2025), spanning November 2020–March 2023 with maximum IVT of approximately 498–1095 kg $m^{-1}$ $s^{-1}$.

2) METRICS

Given an observation field $O \in I \times G$ and a forecast field $F \in I \times G$, where $I$ denotes the number of initial conditions and $G$ the number of grid points within the validation domain, model performance is assessed using RMSE, CC, the ETS, RAPSD, FSS, POD, FAR, and CSI, defined below.

*CIs:* 90% confidence intervals are constructed using a spatial block bootstrap, which better preserves local spatial correlation than an event-based approach (which produced unstable, overly broad intervals given the limited number of independent events). Block size (33 × 33 grid cells) was set from the spatial decorrelation distance of daily PRISM precipitation (132 km — the distance at which mean spatial correlation across randomly sampled grid-cell pairs falls below 0.70); spatial blocks were resampled with replacement (1,000 iterations), and the 5$^{th}$ and 95$^{th}$ percentiles of the resulting distributions define the 90% CIs, using the same resampled blocks across all forecast systems to preserve paired comparisons.

*RMSE:* RMSE (Eq. (1)) is the square root of the mean squared difference between the forecast field $F$ and the observation field $O$. Lower RMSE values are desirable.

$$RMSE = \sqrt{\frac{\sum_{i=1}^{I}\sum_{g=1}^{G}\left(F_{i,g}-O_{i,g}\right)^2}{I \times G}} \tag{1}$$

*CC:* CC (Eq. (2)) measures the strength of the linear relationship between the forecast $F$ and $O$ computed over all initial conditions $I$ and grid points $G$. Values range from −1 to 1, where 1 indicates a perfect positive linear relationship, −1 indicates a perfect negative linear relationship, and 0 indicates no linear correlation.

$$r = \frac{\sum_{i=1}^{I}\sum_{g=1}^{G}\left(F_{i,g}-\bar{F}\right)\left(O_{i,g}-\bar{O}\right)}{\sqrt{\left[\sum_{i=1}^{I}\sum_{g=1}^{G}\left(F_{i,g}-\bar{F}\right)^2\right]\left[\sum_{i=1}^{I}\sum_{g=1}^{G}\left(O_{i,g}-\bar{O}\right)^2\right]}} \tag{2}$$

*ETS:* ETS (Eq. (3)) assesses a model's ability to detect events exceeding a predefined threshold, using the number of hits ($H$), false alarms ($FA$), misses ($M$), correct negatives ($CN$), and the number of hits expected by chance ($E$; Eq. (4)). ETS is calculated using both a fixed precipitation threshold of 0.25 mm day$^{-1}$ defined by the National Weather Service (NWS) and climatological percentile-based thresholds (P50, P90, P98) derived from wet-day precipitation during November–March 2012–2023.

$$ETS = \frac{H-E}{H+FA+M-E} \quad (3)$$

$$E = \frac{(\mathrm{H+FA})(\mathrm{H+M})}{\mathrm{H+FA+M+CN}} \quad (4)$$

<u>*RAPSD:*</u> RAPSD characterizes the distribution of spatial variance across scales in a precipitation field. Each field is standardized (mean-subtracted, normalized by its spatial standard deviation), missing values replaced with zeros, and transformed via a two-dimensional discrete Fourier transform (Eq. (5)); the resulting power spectrum is radially averaged over concentric annular bins of dimensionless radial Fourier wavenumber ($k$) (Eq. (6)) and converted to spatial wavelength using the grid spacing ($\Delta x$ = 4 km for PRISM; Eq. (7)), then averaged across all samples for each model configuration.

$$P(k_x, k_y) = |\mathcal{F}\{X(x, y)\}|^2 \quad (5)$$

where $X$ (x, y) denotes the normalized spatial field and $F$ represents the two-dimensional Fourier transform.

The radially averaged power spectrum is then computed as Eq. (6):

$$RAPSD(k) = \frac{1}{N_k} \sum_{(k_x, k_y) \in A_k} P\left(k_x, k_y\right) \quad (6)$$

where $A_k$ denotes the annular frequency bin associated with radial frequency $k$, and $N_k$ is the number of Fourier coefficients within that annulus.

The approximate spatial wavelength corresponding to each radial frequency bin is computed as in Eq. (7):

$$\lambda_k = \frac{L}{k} \quad (7)$$

where $L = min\ (n_x, n_y)\ \Delta x$ represents the effective domain length and $\Delta x$ is the horizontal grid spacing.

<u>*FSS*</u>*:* FSS (Eq. (8)) evaluates spatial skill by comparing the fractional coverage of threshold-exceeding events within a neighborhood between forecast and observation fields, accounting for small spatial displacement errors. FSS is computed as a function of neighborhood size (fixed 20 mm day$^{-1}$ threshold) and as a function of threshold (fixed 13 × 13 grid-point, ≈50 × 50 km neighborhood), and ranges from 0 (no skill) to 1 (perfect agreement).

$$FSS = 1 - \frac{\sum_{i=1}^{I}\sum_{g=1}^{G}\left(P_{f,i,g} - P_{o,i,g}\right)^2}{\sum_{i=1}^{I}\sum_{g=1}^{G}P_{f,i,g}^2 + \sum_{i=1}^{I}\sum_{g=1}^{G}P_{o,i,g}^2} \quad (8)$$

_POD, FAR, CSI:_ POD (Eq. (9)), FAR (Eq. (10)), and CSI (Eq. (11)) are computed from the number of hits (*H*), false alarms (FA), and misses (*M*). POD measures the fraction of observed events correctly forecast (0–1, higher is better); FAR measures the fraction of forecast events that did not occur (0–1, lower is better); CSI combines both to assess overall detection skill (0–1, higher is better). IVT events use thresholds of 250, 500, and 750 kg m$^{-1}$ s$^{-1}$ (moderate, strong, and extreme AR conditions, respectively); gauge-based precipitation events use each station's own 95$^{th}$-percentile threshold from its 2010–2023 climatology.

$$POD = \frac{H}{H+M} \quad (9)$$

$$FAR = \frac{FA}{H+FA} \quad (10)$$

$$CSI = \frac{H}{H+FA+M} \quad (11)$$

## 3. Results

### *a. Verification Against Gridded Observations*

Fig. 2 compares 4-day precipitation forecasts from West-WRF 3-km, IFS 9-km, West-WRF AI 6-km, and West-WRF AI 2-km against Parameter-elevation Regressions on Independent Slopes Model (PRISM) 4-km observations for the 2021–2022 and 2022–2023 winters (2020–2021 is excluded due to its smaller common domain; Fig. S6), focusing on spatial distribution, frequency, intensity, and AR-related precipitation.

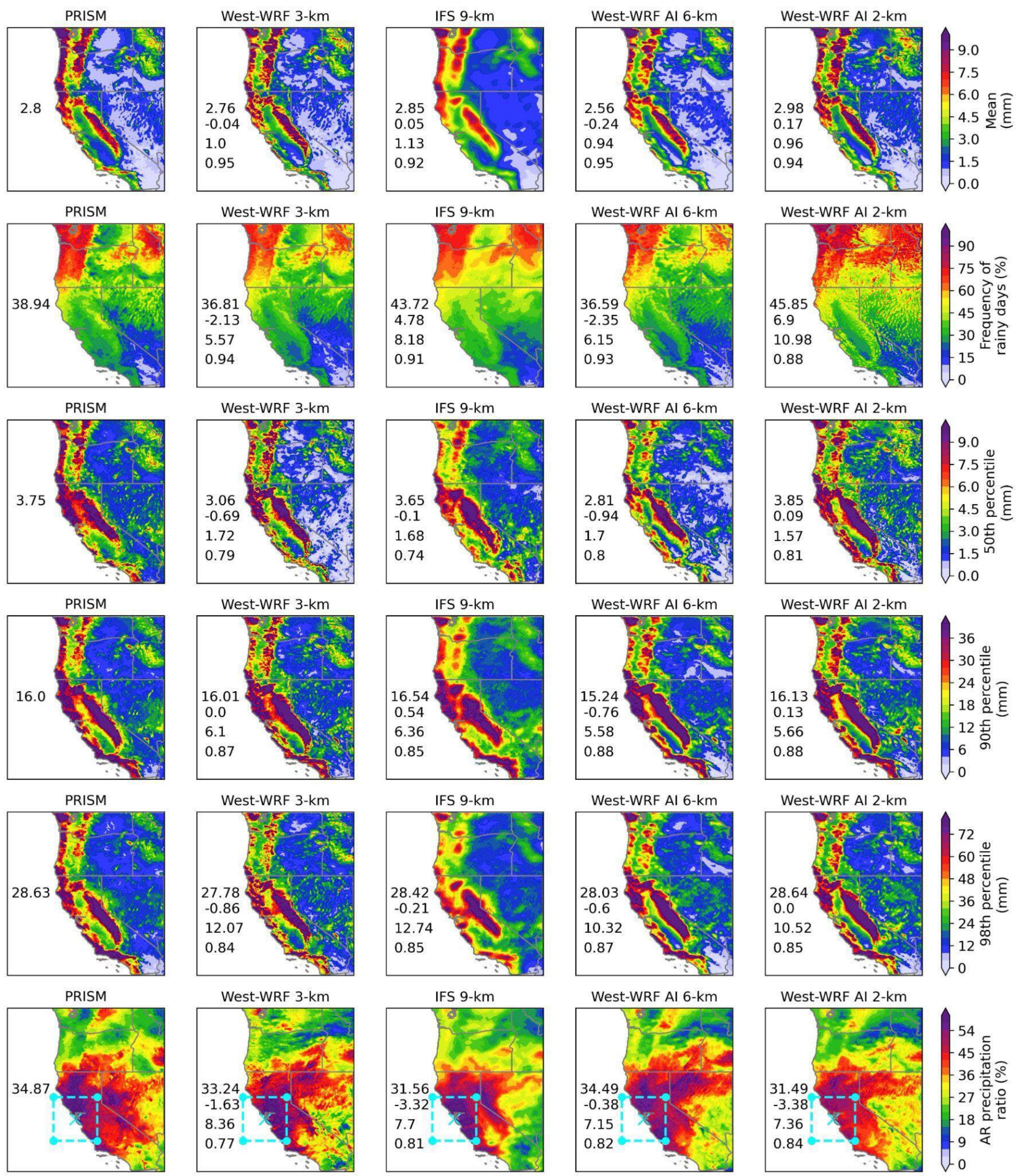

**Fig. 2.** Spatial comparison of observed and forecast precipitation characteristics over the western U.S. during the test period (winters 2021–2022 and 2022–2023) at a 4-day lead time. The columns correspond to PRISM 4-km observations and forecasts from West-WRF 3-km, IFS 9-km, West-WRF AI 6-km, and West-WRF AI 2-km. Rows from top to bottom show daily mean precipitation, rainy-day frequency (> 0.25 mm day$^{-1}$), the 50$^{th}$-, 90$^{th}$-, and 98$^{th}$-percentile precipitation on rainy days, and the AR precipitation ratio. The numbers shown within each forecast panel from top to bottom indicate the spatial mean, bias, root-mean-square error (RMSE), and correlation coefficient (CC) relative to PRISM 4-km observations. Winter 2020–2021 is excluded because the common domain among the models is limited during that season (Fig. S6). The cyan box in the bottom row

indicates the reference region near San Francisco used to identify atmospheric river (AR) days and extends approximately 200 km north, south, east, and west from the center point.

Observations show wetter conditions along the Pacific coast and Sierra Nevada and drier conditions over interior Nevada, Utah, and the desert Southwest, reflecting the strong influence of terrain on both precipitation frequency and intensity.

At this lead time, PRISM's domain-mean daily precipitation is 2.8 mm day$^{-1}$, and all four forecasts reproduce the main coastal and mountain pattern, differing mainly in magnitude and spatial detail. West-WRF 3-km reproduces the PRISM spatial structure, including narrow coastal bands and terrain gradients, with a small negative bias (−0.04 mm day$^{-1}$). IFS 9-km, limited by its coarser resolution, produces smoother, broader patterns with reduced spatial variability; its domain mean (2.85 mm day$^{-1}$) is close to PRISM (bias +0.05 mm day$^{-1}$), but it has the largest root-mean-square error (RMSE) (1.13 mm day$^{-1}$) and lowest spatial correlation (correlation coefficient (CC) = 0.92) of the four forecasts, consistent with smoothing and displacement of precipitation features. West-WRF AI 6-km recovers finer-scale features and sharper terrain gradients than IFS 9-km, with 16.8% lower RMSE, a negative bias of −0.24 mm day$^{-1}$, and an 8.6% underestimation of domain-mean precipitation. West-WRF AI 2-km captures the strongest gradients and most localized features, particularly along the California coastal ranges and Sierra Nevada's western slopes, with a bias of +0.17 mm day$^{-1}$ and RMSE of 0.96 mm day$^{-1}$ — 4.0% lower than West-WRF 3-km, 15.0% lower than IFS 9-km, and essentially on par with West-WRF AI 6-km (0.02 mm day$^{-1}$ higher).

The second row of Fig. 2 shows frequent precipitation over the Pacific Northwest and mountainous coastal regions and drier conditions inland and across the Southwest. Relative to PRISM (38.94%), West-WRF 3-km and West-WRF AI 6-km underestimate rainy-day frequency by approximately 5.5% and 6.0%, respectively, while IFS 9-km and West-WRF AI 2-km overestimate it by approximately 12.3% and 17.7%, respectively.

The 50$^{th}$, 90$^{th}$, and 98$^{th}$ percentiles of rainy-day precipitation (> 0.25 mm day$^{-1}$) represent moderate, heavy, and extreme intensity. At the 50$^{th}$ percentile, all models reproduce the general spatial distribution, though IFS 9-km is smoother than PRISM; relative to PRISM (3.75 mm), West-WRF 3-km, IFS 9-km, and West-WRF AI 6-km underestimate it by approximately 18.4%, 2.7%, and 25.1%, respectively, while West-WRF AI 2-km overestimates it by approximately 2.4% and has the smallest absolute bias and highest CC.

At the 90$^{th}$ and 98$^{th}$ percentiles, differences become more pronounced: IFS 9-km has among the highest RMSE and lowest spatial correlation at both; West-WRF AI 2-km provides the closest agreement with the PRISM domain mean (biases of 0.00–0.13 mm) and reduces RMSE by approximately 7.2–12.8% relative to West-WRF 3-km and 11.0–17.4% relative to IFS 9-km; West-WRF AI 6-km underestimates the domain mean by approximately 2.1–4.8% at these percentiles, while West-WRF AI 2-km reduces the absolute relative bias to approximately 0.04–0.8%, though with slightly higher RMSE than West-WRF AI 6-km.

The bottom row of Fig. 2 shows the fraction of total precipitation associated with AR events. PRISM indicates strong AR contributions along the California coast and nearby mountains (domain mean = 34.87%). West-WRF 3-km underestimates this ratio by approximately 4.7% and IFS 9-km by approximately 9.5%, with smoother, broader patterns; both AI forecasts have lower RMSE and higher spatial correlation than West-WRF 3-km and IFS 9-km. West-WRF AI 6-km provides the closest domain-mean estimate (underestimation of approximately 1.1%) and the lowest RMSE, while West-WRF AI 2-km achieves the highest spatial correlation with PRISM but underestimates the domain mean by approximately 9.7%.

Fig. 3 validates 24-hour accumulated precipitation forecasts against PRISM 4-km during winters 2020–2023 as a function of lead time (1–4 days). RMSE for total and AR-related daily precipitation increases with lead time for all models (panels a, c). West-WRF 3-km consistently produces the largest RMSE, increasing from approximately 4.1 mm on day 1 to 5.4 mm on day 4; the AI forecasts maintain lower RMSE than West-WRF 3-km and IFS 9-km, particularly at longer lead times. For AR-related precipitation, West-WRF 3-km again has the highest RMSE (10.8 to 17.5 mm), while West-WRF AI 6-km and AI 2-km maintain lower errors; IFS 9-km starts low but its error increases rapidly, exceeding both AI forecasts by day 4. CC decreases with lead time for all models (panel b). West-WRF AI 6-km maintains the highest correlation throughout. West-WRF AI 2-km performs comparably to IFS 9-km on day 1 but consistently exceeds it from day 2 onward and outperforms West-WRF 3-km at 3 days or longer, with the gap widening at longer lead times.

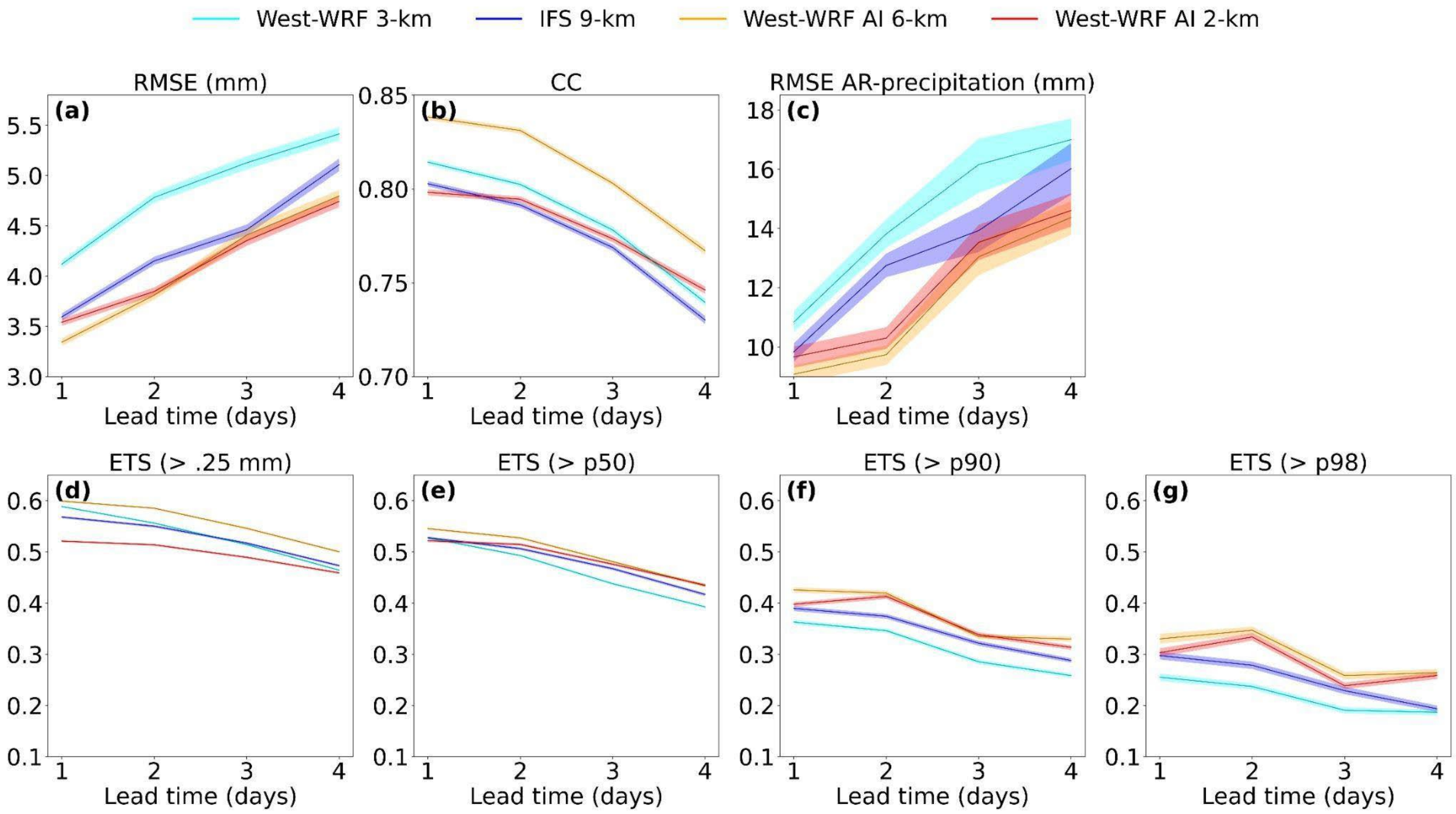

**Fig. 3.** Validation of precipitation forecasts at 4-km resolution during winter 2020–2023 for West-WRF 3-km (cyan), IFS 9-km (blue), West-WRF AI 6-km (orange), and West-WRF AI 2-km (red) as a function of forecast lead time from 1 to 4 days. Panels show (a) RMSE, (b) CC, (c) RMSE for AR-related precipitation events, and equitable threat score (ETS) computed pixel-wise for thresholds of (d) 0.25 mm day$^{-1}$, (e) rainy-day 50$^{th}$ percentile (p50), (f) rainy-day 90$^{th}$ percentile (p90), and (g) rainy-day 98$^{th}$ percentile (p98). All statistics are calculated using 24-hour accumulated precipitation fields relative to PRISM 4-km observations.

Detection skill (equitable threat score (ETS); panels d–g) decreases with increasing threshold and lead time for all models. At the low threshold (>0.25 mm day$^{-1}$), West-WRF AI 6-km achieves the highest ETS throughout; at the median threshold (p50), all models perform comparably. Differences grow at higher thresholds: at p90, West-WRF AI 6-km exceeds West-WRF 3-km by approximately 17–27% and IFS 9-km by 4–15%, while West-WRF AI 2-km improves on West-WRF 3-km by 9–21% and IFS 9-km by 2–10%. At p98, West-WRF AI 6-km exceeds West-WRF 3-km by approximately 29–46% and IFS 9-km by 12–36%, while West-WRF AI 2-km improves by approximately 17–40% and 2–34%, respectively; West-WRF AI 2-km's detection skill remains comparable to West-WRF AI 6-km throughout.

The 90% confidence intervals (CIs) are relatively narrow for most metrics, widening at longer lead times — particularly for AR-related RMSE — consistent with increasing forecast variability with forecast horizons.

Fig. 4 compares precipitation-intensity distributions and spatial variability with PRISM 4-km. All models closely reproduce the PRISM distribution for low-to-moderate amounts across 2–4-day lead times (coefficients of determination ($R^2$) = 0.98–1.00; panels a–c), with larger differences in the upper tail (P99–P100): West-WRF 3-km overestimates the highest PRISM quantiles, IFS 9-km substantially underestimates them, and the AI forecasts fall between the two, reproducing the upper tail better than IFS 9-km while still underestimating the most extreme amounts.

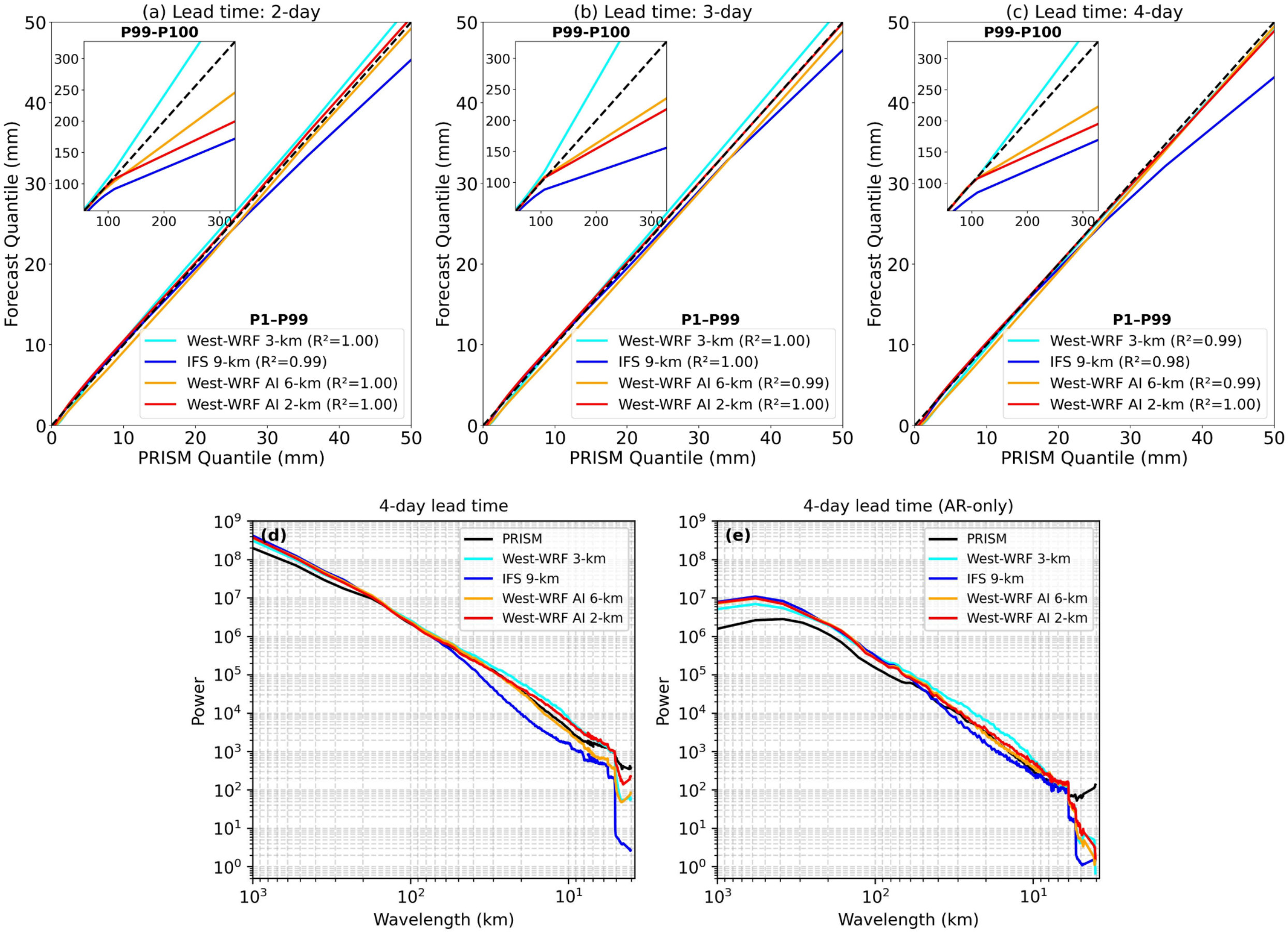

**Fig. 4.** (a)–(c) Quantile–quantile comparisons of precipitation intensity between PRISM and West-WRF 3-km (cyan), IFS 9-km (blue), West-WRF AI 6-km (orange), and West-WRF AI 2-km (red) at lead times of 2, 3, and 4 days, respectively. The main panels show quantiles from P1 to P99, while the inset panels show the upper-tail quantiles from P99 to P100. The dashed black line represents perfect agreement with PRISM. (d)–(e) Radially averaged power spectral density (RAPSD) of the precipitation fields at a 4-day lead time for all events during the test period and for AR events only, respectively. PRISM is shown in black, West-WRF 3-km in cyan, IFS 9-km in blue, West-WRF AI 6-km in orange, and West-WRF AI 2-km in red.

Radially averaged power spectral density (RAPSD) at a 4-day lead time (panels d–e) shows all model spectra following PRISM at large and intermediate scales (>60 km) for all events, with differences emerging at shorter wavelengths: IFS 9-km loses spectral power most rapidly, while West-WRF 3-km and the AI models retain more small-scale variability; at 4–6 km, West-WRF AI 2-km remains closest to PRISM among the four forecasts.

For AR-related precipitation (panel e), all four forecasts exceed PRISM's spectral power above approximately 60 km; the AI forecasts approach the PRISM spectrum at 50–60 km and track it down to about 6 km, whereas IFS 9-km loses power and West-WRF 3-km retains comparatively more; at the smallest resolved scales (<6 km) all models show reduced power relative to PRISM, though West-WRF AI 2-km and West-WRF 3-km retain more than the others.

Fig. 5 compares the fractional skill score (FSS) of the four forecasts at 2- and 4-day lead times as a function of neighborhood window size and precipitation threshold. FSS increases with window size for all forecasts (panels a–b), as larger neighborhoods reduce the penalty for small displacement errors. At the largest window (150 grid points ≈ 600 km), West-WRF 3-km achieves the highest FSS at both lead times, followed by West-WRF AI 2-km, West-WRF AI 6-km, and IFS 9-km; at 2 days, improvements over IFS 9-km are approximately 16.3%, 11.6%, and 3.5%, respectively, and at 4 days, 22.4%, 20.1%, and 14.3%. West-WRF AI 2-km remains particularly close to West-WRF 3-km (absolute FSS differences of 0.026 on day 2 and 0.010 at day 4). FSS decreases with lead time for all models, most for IFS 9-km (approximately 19.3%) and least for West-WRF AI 6-km (10.8%).

FSS also decreases with increasing precipitation threshold for a fixed 13 × 13 grid-point (≈50×50 km) neighborhood (panels c–d). At 2 days, FSS falls from 0.655 to 0.282 for West-WRF 3-km, 0.671 to 0.178 for West-WRF AI 2-km, 0.650 to 0.153 for West-WRF AI 6-km, and 0.663 to 0.065 for IFS 9-km (reductions of approximately 57%, 73%, 77%, and 90%). At 4 days, FSS falls from 0.572 to 0.191, 0.611 to 0.110, 0.570 to 0.106, and 0.582 to 0.018, respectively (reductions of approximately 67%, 82%, 81%, and 97%). IFS 9-km declines most, approaching zero at the highest threshold; West-WRF 3-km retains the greatest fraction of its initial skill; West-WRF AI 2-km ranks among the two best-performing forecasts across both lead times and window sizes and retains substantially more skill than IFS 9-km at higher thresholds.

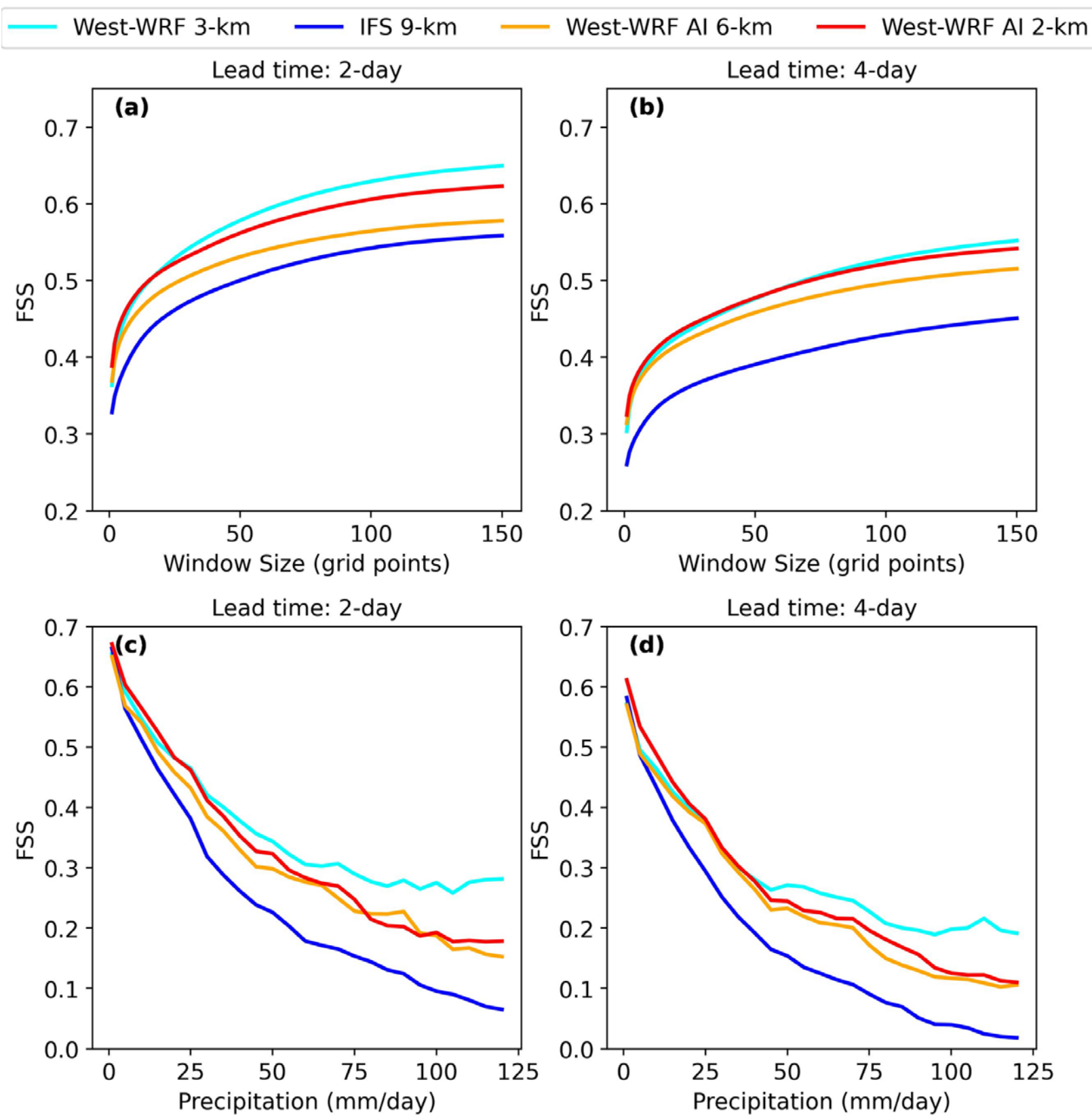

**Fig. 5.** Fractional Skill Score (FSS) for West-WRF 3-km (cyan), IFS 9-km (blue), West-WRF AI 6-km (orange), and West-WRF AI 2-km (red) forecasts at 2- and 4-day lead times during the winters of 2020–2023, evaluated against 4-km, daily PRISM precipitation. Panels (a) and (b) show FSS as a function of neighborhood window size for a fixed precipitation threshold of 20 mm $day^{-1}$. Panels (c) and (d) show FSS as a function of precipitation threshold for a fixed 13 × 13 grid-point neighborhood, corresponding to approximately 50 × 50 km. Higher FSS values indicate greater neighborhood-scale agreement between the forecast and PRISM precipitation fields.

We next present a case study of accumulated precipitation for an AR Scale 2 event on January 8, 2023, forecast from initialization at 00 UTC on January 4 (4-day lead time; Fig. 6; an additional case in Fig. S7). PRISM shows a narrow, intense precipitation band along the California coast and enhanced precipitation over the mountains, with a domain mean of 5.07 mm, compared with 4.55 mm for West-WRF 3-km (−10.3%), 5.32 mm for IFS 9-km (+4.9%), 4.95 mm for West-WRF AI 6-km (−2.4%, smallest bias), and 5.45 mm for West-WRF AI 2-km (+7.5%). West-WRF 3-km and both AI forecasts reproduce finer-scale coastal and mountain features that are less evident in IFS 9-km; West-WRF AI 2-km captures the coastal band and localized maxima with particularly strong spatial agreement to PRISM despite its slight overestimation.

Spatial RMSE is lowest for West-WRF AI 2-km (5.52 mm), followed by West-WRF 3-km (6.79 mm), West-WRF AI 6-km (7.19 mm), and IFS 9-km (8.27 mm) — West-WRF AI 2-km reduces RMSE by 33.3% relative to IFS 9-km, 23.2% relative to West-WRF AI 6-km, and 18.7% relative to West-WRF 3-km. IFS 9-km exhibits broad regional biases along the coast and mountains; West-WRF 3-km captures much of the fine-scale structure but underestimates locally in heavy-precipitation regions, and West-WRF AI 6-km retains some coastal and terrain-related errors.

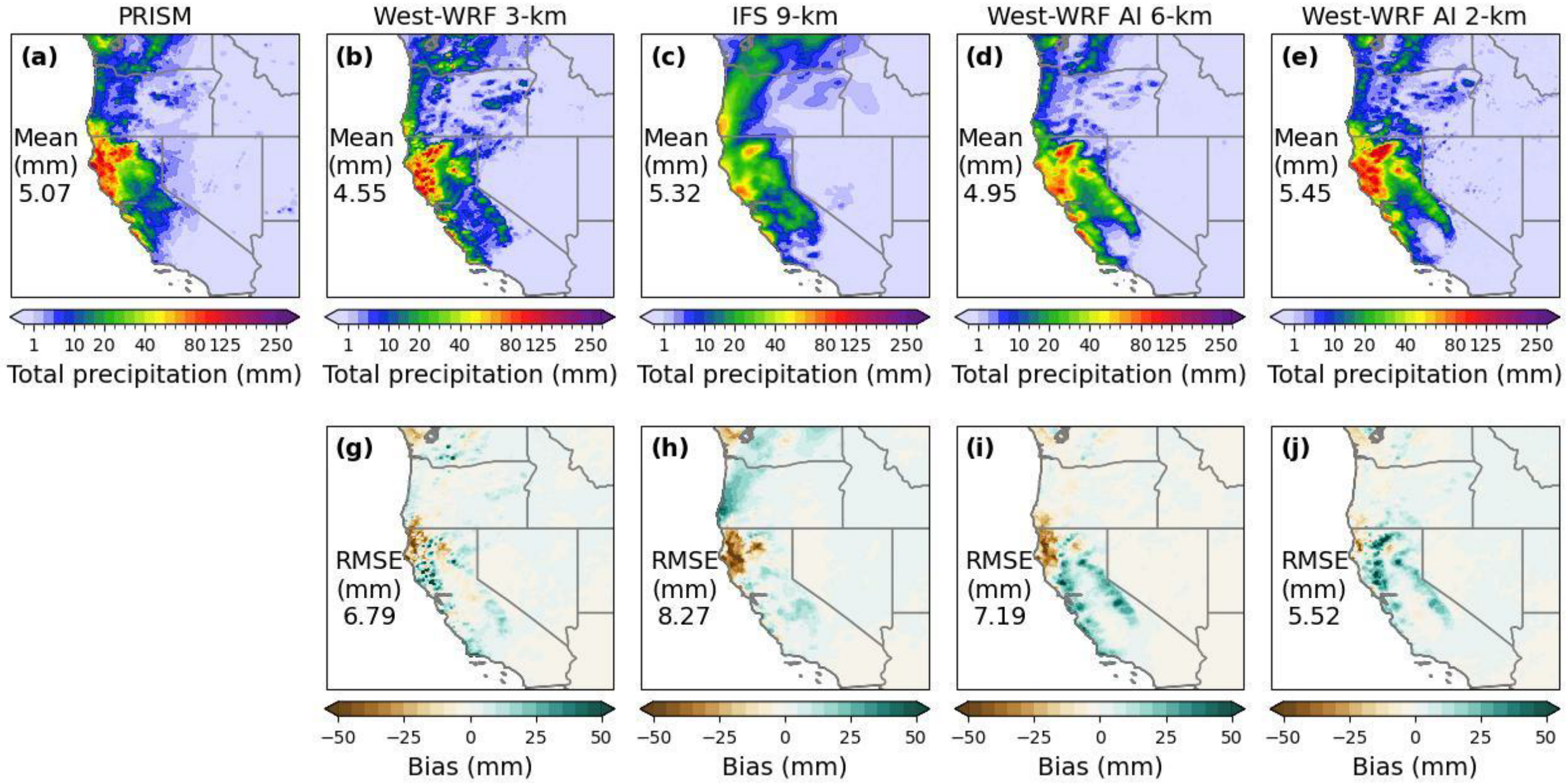

**Fig. 6.** Accumulated precipitation associated with an AR Scale 2 event on January 8, 2023. Forecasts were initialized at 00 UTC on January 4, 2023, corresponding to a 4-day lead time. Panels (a)–(e) show accumulated precipitation from PRISM 4-km observations, West-WRF 3-km, IFS 9-km, West-WRF AI 6-km, and West-WRF AI 2-km, respectively. Domain-mean precipitation is reported within each panel. Panels (g)–(j) show forecast precipitation bias relative to PRISM for West-WRF 3-km, IFS 9-km, West-WRF AI 6-km, and West-WRF AI 2-km, respectively, with the corresponding spatial RMSE reported in each panel. Positive and negative bias values indicate forecast overestimation and underestimation, respectively.

To characterize how refinement from 6 km to 2 km affects AR forecast skill, Fig. 7 presents accumulated precipitation for the AR Scale 3 event of December 31, 2022 (4-day lead time, initialized December 27), comparing PRISM, West-WRF AI 6-km, and West-WRF AI 2-km at native resolution. PRISM shows an intense, spatially confined band along the coast and coastal ranges organized into two high-intensity cores within a broader area of moderate precipitation (domain mean 13.26 mm). West-WRF AI 6-km reproduces the event's general orientation (domain mean 12.84 mm, −3.2%) but smooths the two cores and reduces their peak magnitudes. West-WRF AI 2-km (domain mean 14.55 mm, +9.7%) more clearly captures the localized, high-intensity precipitation smoothed in the 6-km forecast, reproducing both cores with sharper gradients and peak magnitudes closer to PRISM. After resampling both forecasts to the PRISM 4-km grid, West-WRF AI 6-km has a spatial RMSE of 13.38 mm, with underestimation coincident with the most intense core and smaller overestimation elsewhere along the coast; West-WRF AI 2-km has a spatial RMSE of 9.64 mm — approximately 28.0% lower — with smaller, more spatially distributed bias patches of both signs rather than one dominant error region.

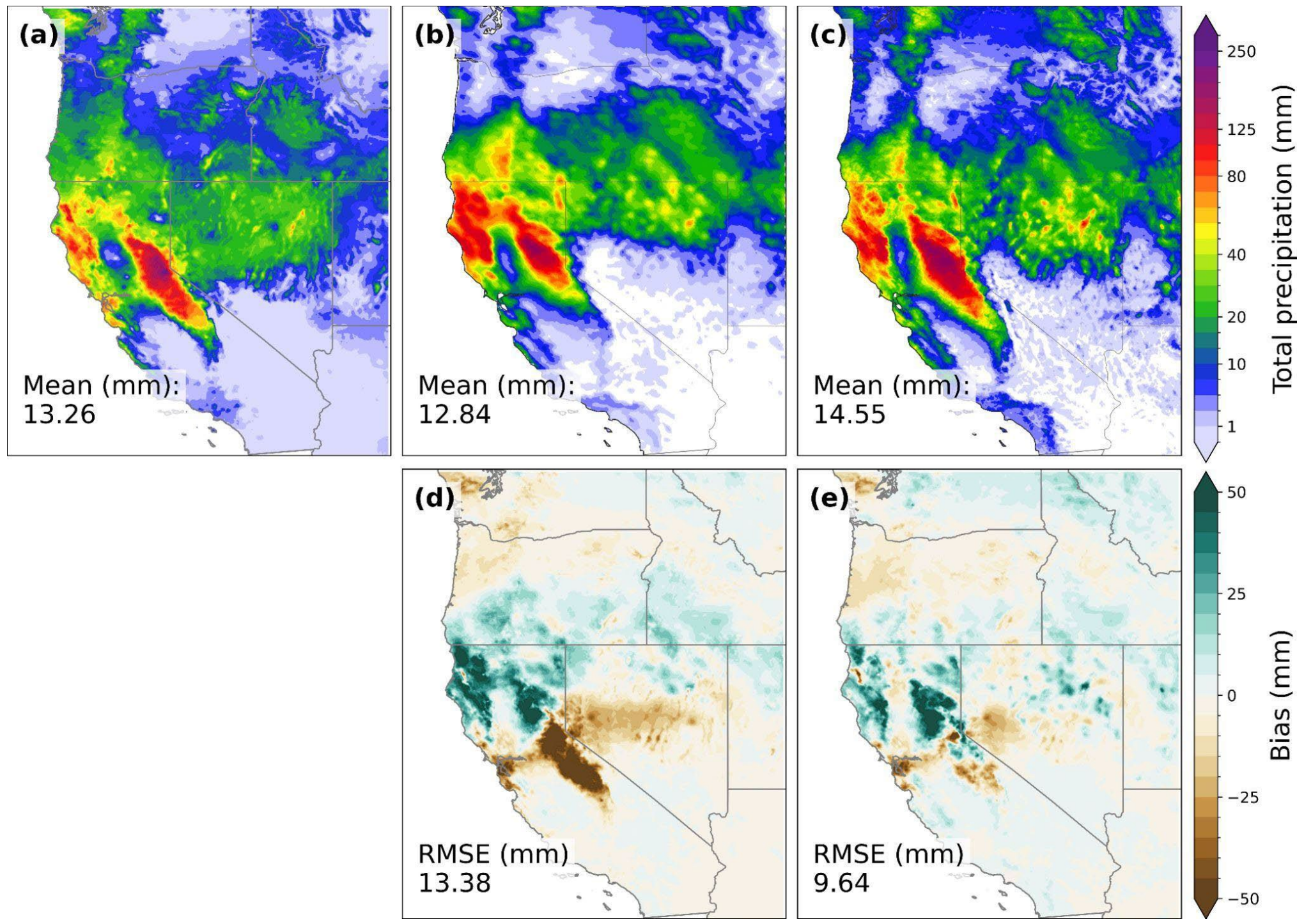

**Fig. 7.** Accumulated precipitation for the AR Scale 3 event of December 31, 2022, from forecasts initialized at 00 UTC on December 27, 2022, corresponding to a 4-day lead time. Panels (a)–(c) show accumulated precipitation (mm) from PRISM 4-km observations, West-WRF AI 6-km, and West-WRF AI 2-km, respectively, each shown at native grid resolution, with domain-mean precipitation reported in each panel. Panels (d)–(e) show forecast precipitation bias relative to PRISM for West-WRF AI 6-km and West-WRF AI 2-km, respectively; forecasts were resampled to the PRISM 4-km grid prior to computing bias and the corresponding spatial RMSE, which is reported in each panel. Positive and negative bias values indicate forecast overestimation and underestimation, respectively.

Results using the higher-resolution PRISM 800-m dataset were consistent with the 4-km evaluation and are provided in the Supplemental Material (Figs. S8–S9).

*b. Verification Against Gauge-Based Observations*

Fig. 8 verifies 6-hour accumulated precipitation forecasts against 526 gauges for lead times of 6–120 hours during winters 2020–2023. RMSE increases and CC decreases with lead time, while mean bias stays small relative to RMSE throughout (Fig. 8(a)–(c)). Forecast errors range from 1.67–1.93 mm during the first 24 hours (Fig. 8(a)). At 72 hours, RMSE is 2.54 mm for West-WRF AI 2-km and 2.52 mm for West-WRF AI 6-km, versus 2.79 mm for IFS 9-km and 3.03 mm for West-WRF 3-km — the AI forecasts reduce RMSE by approximately 9–10% relative to IFS 9-km and 16–17% relative to West-WRF 3-km. At 120 hours, RMSE reaches 3.03, 3.07, 3.09, and 3.76 mm, respectively; West-WRF AI 2-km reduces RMSE by 19.4% relative to West-WRF 3-km and approximately 2% relative to IFS 9-km, while West-WRF AI 6-km reduces RMSE by 18.3% relative to West-WRF 3-km but performs similarly to IFS 9-km.

All four forecasts maintain CC of 0.81–0.87 during the first 24 hours (Fig. 8(b)). At 72 hours, CC is 0.72 for West-WRF AI 2-km and 0.71 for West-WRF AI 6-km, versus 0.65 for West-WRF 3-km and 0.63 for IFS 9-km. At 120 hours, both AI forecasts retain CC ≈ 0.61, versus 0.53 for IFS 9-km and 0.46 for West-WRF 3-km — a 0.05–0.08 advantage over IFS 9-km and 0.13–0.15 over West-WRF 3-km. Mean bias remains small (−0.09 to 0.18 mm) and fluctuates around zero throughout the forecast period (Fig. 8(c)). Overall, the two AI forecasts remain competitive with or outperform the dynamical forecasts across much of the five-day period, with their advantage growing at intermediate and longer lead times.

Fig. 8(d)–(f) present categorical verification for 6-hour precipitation events exceeding each gauge's station-specific $95^{th}$-percentile threshold. Probability of detection (POD) and critical success index (CSI) decrease with lead time while false-alarm ratio (FAR) increases; West-WRF 3-km and the two AI forecasts achieve comparable POD, with the best-performing regional model varying by lead time. At 24 hours, POD is highest for West-WRF AI 2-km (0.516), followed by West-WRF 3-km (0.501), West-WRF AI 6-km (0.478), and IFS 9-km (0.386).

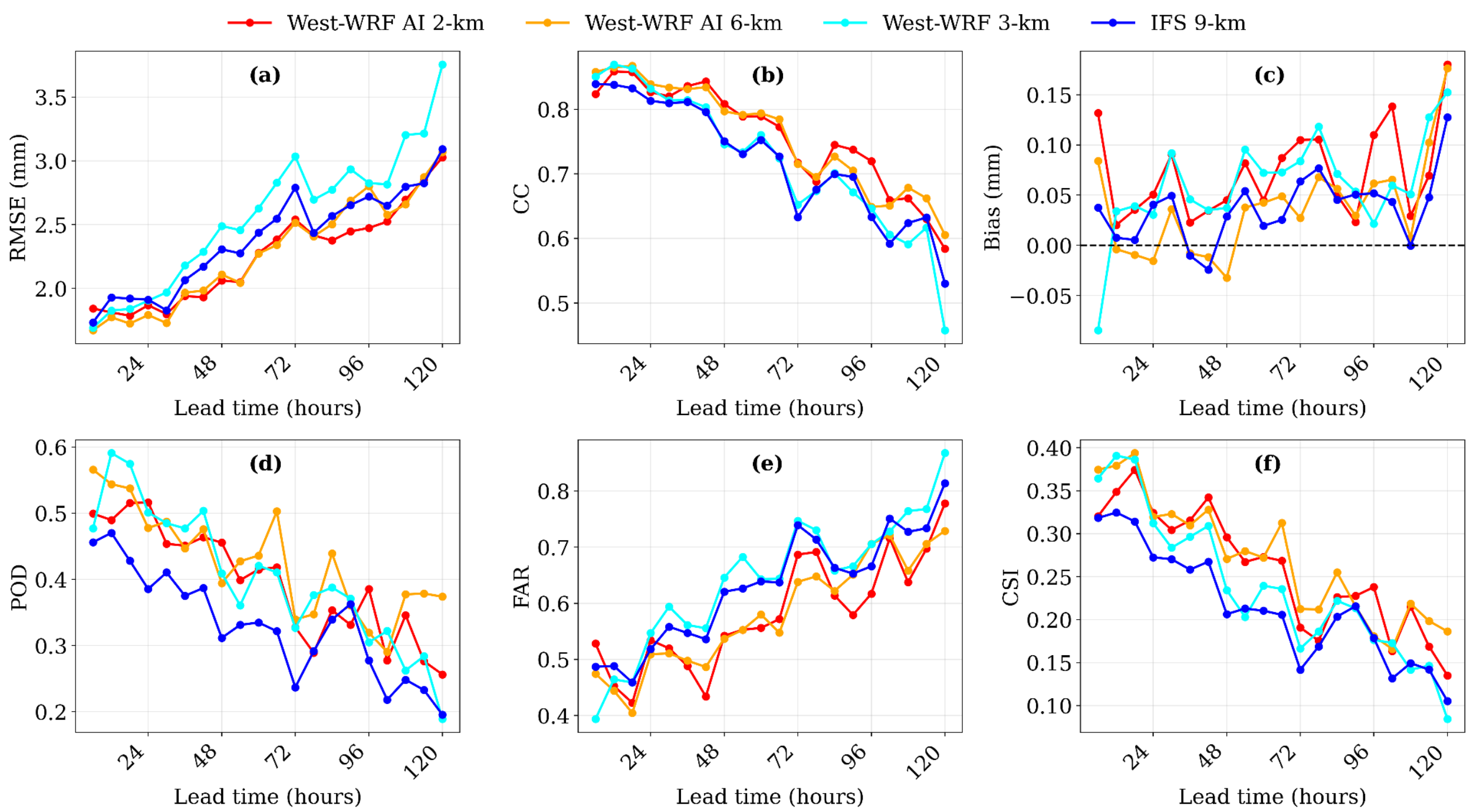

**Fig. 8.** Verification of 6-hour accumulated precipitation forecasts against gauge observations for lead times from 6 to 120 hours during winters 2020–2023. Panels show (a) RMSE (mm), (b) CC, and (c) mean bias (forecast minus observation; mm) for West-WRF AI 2-km (red), West-WRF AI 6-km (orange), West-WRF 3-km (cyan), and IFS 9-km (blue). Positive and negative bias values indicate precipitation overestimation and underestimation, respectively. Panels (d)–(f) show probability of detection (POD), false-alarm ratio (FAR), and critical success index (CSI), respectively for 6-hour precipitation events exceeding the station-specific 95th-percentile precipitation threshold. Higher POD and CSI and lower FAR indicate better categorical forecast performance. Metrics and thresholds were calculated using 526 valid gauges, with thresholds calculated separately for each gauge.

FAR is similar across models early on (0.51–0.55 at 24 hours) and increases with lead time (Fig. 8(e)). The two AI forecasts trade off: West-WRF AI 2-km has lower FAR at 42, 60, and 84–114 hours, while West-WRF AI 6-km performs better at other lead times, including 72 and 120 hours. The dynamical models have the highest FAR at most lead times. CSI, which combines detection and false alarms, is 0.324 for West-WRF AI 2-km, 0.320 for West-WRF AI 6-km, 0.312 for West-WRF 3-km, and 0.273 for IFS 9-km at 24 hours (17–19% higher for the AI forecasts). At 120 hours, CSI is 0.186 for West-WRF AI 6-km, 0.135 for West-WRF AI 2-km, 0.105 for IFS 9-km, and 0.084 for West-WRF 3-km — from 24 to 120 hours, West-WRF AI 6-km improves on IFS 9-km and West-WRF 3-km by approximately 77% and 121%, respectively, while West-WRF AI 2-km improves by approximately 29% and 61% (Fig. 8(f)).

### *c. AR Reconnaissance Dropsonde IVT*

Fig. 9 verifies IVT forecasts against AR Reconnaissance dropsonde observations (2,360 samples) for West-WRF AI 2-km/31-km, West-WRF AI 6-km, and West-WRF 9-km at lead times of 24–120 hours. Skill decreases with lead time for all three systems, as indicated by increasing RMSE and decreasing CC. At 24 hours, the three systems perform similarly (RMSE = 124.37, 124.94, and 128.43 kg $m^{-1}$ $s^{-1}$; CC = 0.85, 0.85, and 0.84), with West-WRF AI 2-km/31-km reducing RMSE by approximately 3.2% relative to West-WRF 9-km and matching West-WRF AI 6-km. The gap widens steadily through 72 and 96 hours, reaching its largest separation at 120 hours, where RMSE is 183.30, 194.16, and 199.40 kg $m^{-1}$ $s^{-1}$ (5.6% and 8.1% lower than West-WRF AI 6-km and West-WRF 9-km, respectively) and CC is 0.67, 0.64, and 0.62. West-WRF AI 2-km/31-km consistently produces the lowest RMSE and highest CC across all lead times, with West-WRF AI 6-km performing between it and West-WRF 9-km; differences are small at short lead times but grow at 96 and 120 hours, indicating slower skill degradation for West-WRF AI 2-km/31-km.

Fig. 9(c)–(e) show categorical performance for IVT thresholds of 250, 500, and 750 kg $m^{-1}$ $s^{-1}$ (moderate, strong, and extreme AR conditions), which degrades with lead time for all configurations. At the 250 kg $m^{-1}$ $s^{-1}$ threshold, POD starts near 0.88 for all models at 24 hours and falls to approximately 0.75–0.76 for the AI configurations and 0.71 for West-WRF 9-km at 120 hours; at 500 kg $m^{-1}$ $s^{-1}$, POD falls from approximately 0.81–0.82 to 0.55–0.58; the largest decline is at 750 kg $m^{-1}$ $s^{-1}$, from approximately 0.75–0.76 to 0.39–0.40. The two AI configurations maintain broadly comparable POD across most lead times, while West-WRF 9-km declines more, particularly at the higher thresholds.

FAR increases with lead time for all models: at 250 kg $m^{-1}$ $s^{-1}$, from approximately 0.06–0.08 to 0.16–0.18; at 500 kg $m^{-1}$ $s^{-1}$, from approximately 0.12–0.15 to 0.32–0.37; at 750 kg $m^{-1}$ $s^{-1}$, the largest increase, from approximately 0.21 at 24 hours to about 0.49 for West-WRF AI 2-km/31-km and approximately 0.56 for the other two configurations at 120 hours—a 13% reduction for West-WRF AI 2-km/31-km relative to the others.

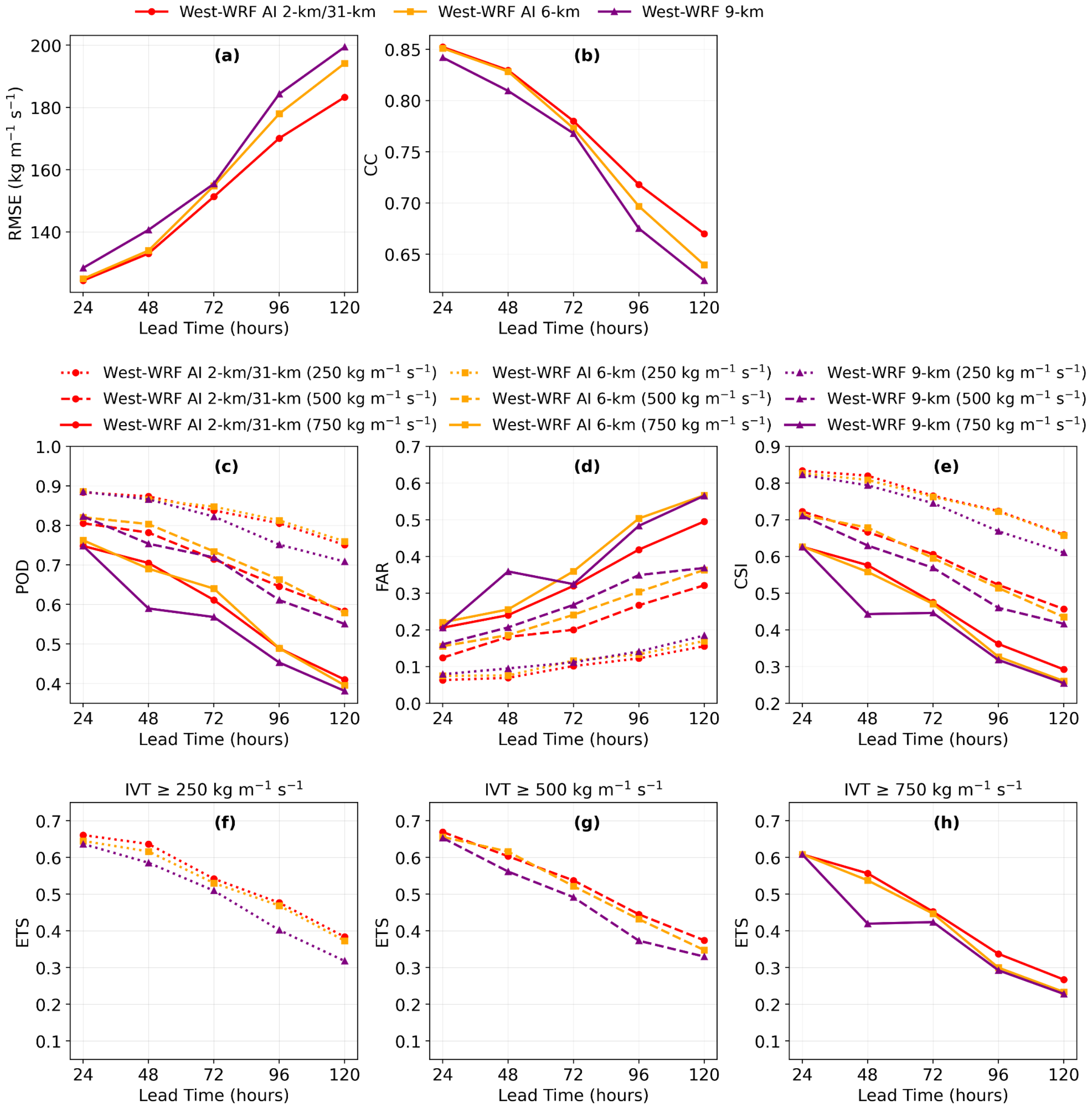

**Fig. 9.** Verification of integrated vapor transport (IVT; kg m⁻¹ s⁻¹) forecasts against AR Reconnaissance dropsonde observations at forecast lead times from 24 to 120 hours for West-WRF AI 2-km/31-km (red), West-WRF AI 6-km (orange), and West-WRF 9-km (purple), using 2,360 dropsonde samples at each lead time. Panels (a) and (b) show RMSE and CC, respectively. Panels (c)–(e) show POD, FAR, and CSI, respectively, for IVT thresholds of 250 kg m⁻¹ s⁻¹ (dotted), 500 kg m⁻¹ s⁻¹ (dashed), and 750 kg m⁻¹ s⁻¹ (solid). Panels (f)–(h) show ETS for IVT thresholds of ≥250, ≥500, and ≥750 kg m⁻¹ s⁻¹, respectively. Lower RMSE and FAR and higher CC, POD, CSI, and ETS indicate better forecast performance. Positive ETS values indicate forecast skill above that expected from random chance.

CSI decreases with lead time and threshold (Fig. 9(e)). At 250 kg $m^{-1}$ $s^{-1}$, CSI falls from approximately 0.81–0.82 to about 0.66 for the AI models and 0.60 for West-WRF 9-km (approximately 10% higher for the AI configurations at 120 hours). At 500 kg $m^{-1}$ $s^{-1}$, CSI falls to approximately 0.43–0.45 for the AI models and 0.41 for West-WRF 9-km. At 750 kg $m^{-1}$ $s^{-1}$, CSI falls from approximately 0.62 to 0.29 for AI 2-km/31-km, 0.26 for AI 6-km, and 0.25 for West-WRF 9-km—AI 2-km/31-km improves on AI 6-km by approximately 12% and on West-WRF 9-km by 16% at this threshold and lead time.

ETS decreases with lead time and threshold (Fig. 9(f)–(h)). At 250 kg $m^{-1}$ $s^{-1}$, ETS falls from approximately 0.64–0.66 at 24 hours to 0.38–0.39 for the AI models and 0.31 for West-WRF 9-km at 120 hours (23–26% higher for the AI configurations). At 500 kg $m^{-1}$ $s^{-1}$, ETS falls to approximately 0.35–0.38 for the AI models and 0.33 for West-WRF 9-km (AI 2-km/31-km approximately 15% higher). The reduction is most pronounced at 750 kg $m^{-1}$ $s^{-1}$: all models start near 0.61 at 24 hours, falling to approximately 0.27 for AI 2-km/31-km and 0.23 for both AI 6-km and West-WRF 9-km by 120 hours—a 17% improvement for AI 2-km/31-km. All models retain positive ETS throughout; West-WRF 9-km consistently shows the lowest values, particularly for the strongest events and longer lead times.

## 4. Discussion and Conclusions

This study evaluates how refinement from West-WRF AI 6-km (Baño-Medina et al. 2025) to 2-km affects AR-related IVT and orographic-precipitation forecasts while retaining approximately 31-km global context (Baño-Medina et al. 2025). The objective is not to introduce new architecture but to assess resolution sensitivity. To our knowledge, West-WRF AI 2-km is among the first regional AI systems evaluated specifically for AR prediction at this spatial detail, distinguishing it from regional AI efforts not focused on ARs or IVT (the Norwegian Meteorological Institute’s (MET Norway) Bris (Nipen et al. 2025); Adamov et al.'s Alpine LAMs (Adamov et al. 2025)).

The 2-km configuration’s advantage is concentrated, not uniform, and is clearest over complex terrain and along the coast, where sharp, localized gradients remain smoothed at 6-km resolution. Differences between the two AI configurations are small near the median of the precipitation distribution but grow at the upper percentiles, where West-WRF AI 2-km more closely reproduces

PRISM — indicating a favorable balance between realistic precipitation intensity and fine-scale spatial fidelity, without the stronger upper-tail overestimation seen in West-WRF 3-km.

The December 31, 2022, case study illustrates this clearly: West-WRF AI 2-km reproduces both intense precipitation cores with sharper gradients and a substantially lower spatial RMSE than West-WRF AI 6-km (9.64 versus 13.38 mm), which instead smooths and displaces the most intense core, consistent with improved representation of localized precipitation structure in regions influenced by complex terrain and coastal processes. Across the broader model comparison, West-WRF AI 2-km also improves on the smoother IFS 9-km forecasts, which struggle with localized extremes, narrow coastal bands, and terrain-induced maxima, and remains competitive with West-WRF 3-km, which is itself highly competitive for some neighborhood-based and extreme-precipitation metrics. This picture is supported by the independent gauge verification, where both AI configurations remain competitive with or outperform the dynamical forecasts across much of the five-day period, with the advantage growing at intermediate and longer lead times, even though neither AI configuration is uniformly best for station-specific extremes at every lead time.

Independent dropsonde verification further supports the results: West-WRF AI 2-km/31-km shows smaller errors and slower skill degradation, with differences from West-WRF AI 6-km and West-WRF 9-km increasing at 96–120-hour lead times. However, the AI configurations differ in both regional resolution and outer-domain configuration; therefore, controlled experiments are needed to isolate the effect of inner-grid resolution.

At short IVT lead times (Fig. 9), for domain-mean and lower-percentile precipitation intensity (Fig. 3), and for neighborhood-based skill at large spatial windows (Fig. 5), West-WRF AI 2-km performs comparably to — rather than clearly better than — West-WRF AI 6-km. This does not indicate that higher resolution provides no benefit; rather, refining the grid to 2 km preserves large-scale and short-range skill while adding spatial detail, and the benefits become evident where localized terrain effects and extreme precipitation processes play a greater role.

For FSS at large neighborhoods (Fig. 5) and categorical metrics for station-specific extremes at 120 hours (Fig. 8), the smoother West-WRF AI 6-km forecast sometimes performs better. Large-neighborhood FSS emphasizes broad-scale coverage, limiting the measurable benefit of 2-km detail. This also explains why West-WRF 3-km can have higher pointwise RMSE but higher FSS: neighborhood verification tolerates small displacements, whereas pointwise metrics impose a double penalty when a localized feature is slightly misplaced. High-resolution forecasts are also more sensitive to such displacement errors: an intense, localized feature predicted slightly away from its observed location can produce both a miss and a false alarm (the double-penalty problem), for which a smoother forecast is penalized less. This tendency is compounded because training with a mean-squared-error (MSE) loss favors spatially averaged predictions under positional uncertainty, an effect that grows with lead time as the sharper features from West-WRF AI 2-km become more susceptible to displacement.

The December 31, 2022, case again illustrates the tradeoff: although West-WRF AI 6-km has the smaller domain-mean bias, West-WRF AI 2-km represents the localized cores more clearly and reduces spatial RMSE — no single score fully captures the value of increased resolution alone. The station-based reversal at 120 hours should also be interpreted cautiously: because POD, FAR, and CSI are computed for rare, station-specific 95$^{th}$-percentile events, the effective sample size is limited, and additional winters would help determine whether this reflects a systematic lead-time dependence or sampling variability.

West-WRF AI 2-km is evaluated deterministically to isolate the effect of resolution refinement, consistent with its 6-km predecessor (Baño-Medina et al. 2025); extending to ensemble and probabilistic prediction is a logical next step.

The model was regionally fine-tuned using eight years of CW3E 2-km reanalysis, building on pretraining with approximately 40 years of global ERA5 data, with winters 2020–2023 reserved for independent evaluation; the eight-year regional fine-tuning record reflects the availability of the high-resolution reanalysis rather than a limitation of the modeling framework, and the resulting skill suggests it is sufficient for the present evaluation, though additional years would improve representation of rare events. Because the 2-km regional refinement and regional fine-tuning are restricted to the CW3E reanalysis domain, offshore verification of the high-resolution component is geographically limited, and because the model was trained on ERA5 while real-time forecasting requires operational-analysis initialization, this training–operational mismatch needs further evaluation; the use of reanalysis to initialize forecasts may itself provide an advantage over dynamical forecasts initialized from operational analyses, so fine-tuning and evaluating West-WRF AI 2-km with operational ECMWF IFS or NCEP GFS analyses would give a more representative assessment of real-time performance.

West-WRF AI 2-km produces a higher rainy-day frequency than PRISM despite a small domain-mean intensity bias, echoing an overestimation of wet-day occurrence reported for the 6-km predecessor that may partly reflect biases inherited from ERA5 (Baño-Medina et al. 2025). Evaluating rainy-day frequency at multiple thresholds and comparing wet-day occurrence across ERA5, CW3E reanalysis, PRISM, and both AI configurations would help determine whether this behavior originates in the global inputs, the regional training target, or the refinement stage.

Although MSE-based training often over-smooths, training on high-resolution CW3E 2-km fields appears to mitigate this and preserve finer spatial features; future probabilistic extensions — continuous ranked probability score (CRPS)-based loss functions, diffusion models, or perturbed initial conditions/parameters — could better characterize forecast uncertainty and the upper tail of the precipitation distribution, and future work could also examine sensitivity to training design, such as withholding high-impact ARs, structure-aware or multiscale loss functions, and latent-grid or -channel capacity, though all would need direct evaluation for this system. This study's empirical, application-focused framing — holding the AI architecture fixed while evaluating what resolution refinement gains, where, and under what conditions — is intentional, as establishing when higher resolution improves AR and precipitation forecasts is a necessary step toward guiding future architectural development.

West-WRF AI 2-km does not outperform its 6-km predecessor under every metric or condition: the two remain comparable for domain-mean quantities, broader spatial scales, lower precipitation thresholds, and short-lead IVT, while the clearest benefits of the 2-km configuration occur for localized, terrain-sensitive, extreme AR-related precipitation and intense IVT at longer lead times, with no evident reduction in large-scale skill. The value of increased resolution therefore depends on the forecast variable, spatial scale, lead time, verification method, and application. These findings provide a deterministic baseline for future work on operational-analysis initialization, probabilistic ensemble forecasting, uncertainty-aware training, and expansion of the 2-km domain.

*Acknowledgments*

This research is supported by the Office of Naval Research (ONR) (award N000142412731), the California Department of Water Resources Atmospheric River Program Phase V (Grant 4600015671), and the US Army Corps of Engineers (USACE) Forecast Informed Reservoir Operations Phase 3 (USACE W912HZ-24-2-0001).

The authors gratefully acknowledge the computational resources provided by the National Artificial Intelligence Research Resource Pilot (NAIRR Award 240367) and the DeltaAI advanced computing and data resource, which is supported by the National Science Foundation (award NSF-OAC 2320345).

The authors acknowledge California Energy Commission award PIR-19-007 that supported the development of the CW3E 40-year high-resolution reanalysis. Additionally, we would like to thank Drs. Ming Pan and Qian Cao for their assistance with the quality-control process of the precipitation gauge dataset.

The authors declare no conflicts of interest.

*Availability Statement*

The West-WRF AI 2-km model was trained using the Anemoi framework that is publicly available at https://github.com/ecmwf/anemoi

The PRISM analysis data are publicly available from Oregon State University at https://www.prism.oregonstate.edu/. ERA5 and IFS are publicly available datasets readily accessible from the Copernicus climate data store (https://cds.climate.copernicus.eu) and TIGGE (https://apps.ecmwf.int/datasets/data/tigge/levtype=sfc/type=cf/) data archives, respectively. All CW3E datasets are available from Daniel F. Steinhoff upon request (dsteinhoff@ucsd.edu). Precipitation gauge data were collected at https://madis.ncep.noaa.gov/. The raw gauge data sources include the Geostationary Operational Environmental Satellites (GOES) Data Collection Platforms (DCPs) [data received from the National Oceanic and Atmospheric Administration (NOAA) Hydrometeorological Automated Data System (HADS)], Automated Local Evaluation in Real Time (ALERT) system (data collected locally at the county/water agency), the California Department of Water Resources (CA-DWR)/the California Data Exchange Center (CDEC) gauges, SNOTEL data (queried from the Natural Resources Conservation Service (NRCS) database by Meteorological Assimilation Data Ingest System (MADIS)), as well as the Automated Surface/Weather Observing System (ASOS/AWOS) gauges. For the QC process performed by CW3E, the method is based on the Mountain Mapper (MM) approach, which combines gauge observations with PRISM climatology to account for terrain effects, particularly in mountainous regions. For each gauge, precipitation is compared with estimates derived from nearby stations using inverse distance weighting (IDW) and climatological scaling. Observations that deviate significantly from neighboring stations are flagged as failing QC. The quality-controlled hourly observations were then aggregated to 6-hourly totals and compared with the CNRFC 6-hourly QC product, showing good agreement.